\documentstyle[floats,preprint,prb,aps,tighten]
{revtex}
\parskip=\medskipamount
\begin{document}
\input{epsf} 
\draft
\renewcommand{\textfraction}{0.0}
\renewcommand{\dblfloatpagefraction}{0.8}
\renewcommand{\topfraction}{1.0}   
\renewcommand{\bottomfraction}{1.0}   
\renewcommand{\floatpagefraction}{0.8}
\def\ep{\epsilon}
\def\al{\alpha}
\def\be{\beta}
\def\ga{\gamma}
\def\la{\lambda}
\def\th{\theta}
\def\de{\delta}
\def\si{\sigma}
\def\ti{\tilde}
\def\et{\eta}
\def\pa{\partial}
\def\om{\omega}
\def\fr{\frac}
\def\be{\begin{equation}}
\def\ee{\end{equation}}
\def\bea{\begin{eqnarray}}
\def\eea{\end{eqnarray}}
\title{Parametric Simultons in Nonlinear Lattices}
\author{Bambi Hu$^{1,2}$ and Guoxiang Huang$^{1,3}$ }
\address{ 
          $^{1}$Centre for Nonlinear Studies and Department of Physics, Hong Kong
                Baptist University,\\ Hong Kong, China\\ 
          $^{2}$Department of Physics, University of Houston, Houston TX 77204,
               USA \\           
          $^{3}$Department of Physics and Laboratory for Quantum Optics,
                East China Normal University,
                Shanghai 200062,
                China
          }
\maketitle
\begin{abstract}
Parametric simultaneous solitary wave (simulton) excitations are shown possible
in nonlinear lattices. Taking a one-dimensional diatomic
lattice with a cubic potential as an example we consider the nonlinear coupling 
between the upper cutoff mode of acoustic branch\,(as a fundamental wave) 
and the upper cutoff mode of optical branch\,(as a second harmonic wave). 
Based on a quasi-discreteness approach the 
Karamzin-Sukhorukov equations for two slowly varying amplitudes of the 
fundamental and the second harmonic waves in the lattice
are derived when the condition of 
second harmonic generation is satisfied. The lattice simulton solutions 
are explicitly given and the results show that these lattice simultons
can be nonpropagating when the wave vectors of the fundamental wave
and the  second harmonic waves are exactly at $\pi/a$\,(where $a$ is the lattice 
constant) and zero, respectively.
\end{abstract}
%
\section{INTRODUCTION}
Since the pioneering work of  Fermi, Pasta, and Ulam[1]
on the nonlinear dynamics in lattices, the understanding of the dynamical
localization in ordered, spatially extended discrete systems have 
experienced considerable progress. In particular, the lattice solitons,
which are localized nonlinear excitations due to the balance between 
nonlinearity and dispersion of the system, are shown to exist, and many 
important applications are found in transport of energy, proton
contactivity, structural phase transition and associated central-peak
phenomena, etc[2,3]. In recent years, the interest in localized 
excitations in nonlinear lattices has been renewed due to the identification
of a new type of anharmonic localized  modes[4-6]. These modes, called
the intrinsic localized modes\,(ILM's)[4], or discrete breathers[5,6], are
the discrete analog of the lattice envelope\,(or breather) solitons with their 
spatial extension only a few lattice spacing and the vibrating frequency 
above the upper cutoff of phonon spectrum band. The ILM's have been observed in
a number of experiments[7-13]. Recently, much attention has been paid to
the gap solitons in diatomic lattices[14-25]. In linear case, a diatomic lattice
allows two phonon bands. There is a upper cutoff for phonon
frequency and a frequency gap\,(forbidden band) between acoustic and 
optical bands, induced by mass and/or force-constant difference of two different
types of particles. No interaction occurs between phonons and the phonons can not 
propagate in the system when their frequencies are in the gap or 
above the phonon bands. However,
these properties of the phonons change drastically when nonlinearity is
introduced into the system. New types of localized modes, 
especially the gap solitons, may appear as nonlinear
localized excitations with their vibrating frequencies in the band gap. The
gap solitons and ILM's as well as their chaotic motion have been observed
in damped and parametrically excited one-dimensional\,(1D) diatomic 
pendulum lattices[26-28].

On the other hand, in recent years numerous achievements have been made
for optical solitons in nonlinear optical media[29-31]. Besides the 
temporal optical solitons, which is promising for long-distance information
transmission in fiber, spatial optical solitons also attract much attention.
The spatial optical solitons are believed to be the candidates for all-optical
devices, such as optical switches and logic gates, etc[32]. Recently, the
study of optical parametric processes, in particular the second harmonic 
generation\,(SHG), which marked the birth of nonlinear optics, has generated 
a great deal of new interest[33]. It was suggested that it is possible 
to obtain large nonlinear phase shifts by using a cascaded second-order
nonlinearity[34]. In 1974, Karamzin and Sukhorukov\,(KS) recognized that
the cascaded second-order parametric processes may support solitons under
general phase-matching conditions. They derived two coupled nonlinear equations
for the envelopes of the fundamental and second harmonic waves[35]. 
The difference between the KS equations and the envelope equations for usual SHG
is the inclusion of dispersion and/or diffraction, which 
are necessary for short pulses and/or narrow light beams. 
Simultaneous solitons\,(i.\,e. two components are solitons)
are found for the KS equations and these solitons are later termed as the
{\it parametric simultons}[36]. The concept of the simultons has been generalized 
to the nonlinear optical media with periodically varying refractive index[37].
Since the eigenspectrum of linear electromagnetic waves consists of many photonic
bands and the vibrating frequencies of the simultons may be in the gaps between
these bands, the name {\it parametric band-gap simulton} is given by 
Drummond et al[37-39]. Different from the self-trapping mechanism for Keer
solitons, the formation of the simultons is due to the energy transfer
and {\it mutual self-trapping} 
between the fundamental and the second harmonic waves.

On the contrary, the SHG in lattices is much less investigated. Although in the 
standard textbook of solid state physics[40] there exists a simple experimental
description for three phonon processes in solids, it seems that there 
is no detailed theoretical approach to the SHG in nonlinear lattices until 
recently. In a recent paper, Konotop considered theoretically
the SHG in a nonlinear diatomic lattice and obtained some interesting results[41].

In many aspects a nonlinear lattice is similar to a nonlinear, periodic
optical media. The discreteness of lattice results in the symmetry
breaking for continuous translation and makes the property of the 
system periodic, in particular the frequency spectrum of corresponding 
linear wave splits into many bands. It should be stressed that the SHG does not
occur in 1D monatomic lattices\,(see the next section). However, 
a SHG can be realized 
if we consider nonlinear multi-atomic lattices . The reason is that 
in the monatomic lattices, an efficient energy transfer\,(resonance)
between any two modes in the system does not occur. But the situation
is different for the multi-atomic lattices. A
multi-atomic lattice allows many branches of linear dispersion relation, 
and the dispersion relation is periodic with respect to lattice
wave vector. It is just the multiple-value and periodic property
of the dispersion relation makes it possible that the 
phase-matching condition for the SHG, i.\,e. the condition by which the resonance 
between the fundamental and second harmonic waves may occur, 
can be satisfied by selecting the 
wave vectors and the corresponding frequencies from different spectrum 
branches.

Motivated by the study of the optical simultons, in this paper we show that 
lattice simultons are possible in the multiatomic lattices with cubic
nonlinearity\,(different from the case in nonlinear optics, 
here the order of nonlinearity means the order in the Hamiltonian of the system). 
The paper is organized as follows. 
The next section presents our model and an asymptotic
expansion based on a quasi-discreteness approach. 
In section III we solve the KS equations derived in section II
and provide some lattice simulton solutions. A discussion and summary is 
given in the last section.  
%
%
\section{MODEL AND ASYMPTOTIC EXPANSION}
%
\subsection{The model}
As mentioned in the last section, the SHG may in princile
occur in any multi-atomic lattice, but for definiteness and for the sake of 
simplicity we consider here a 1D diatomic lattice with a cubic interaction
potential. The Hamiltonian of the system is given by
\bea
& & H=\sum_n\left [ \fr{1}{2}m \left (\fr{dv_n}{dt}\right )^2
               +\fr{1}{2}M \left (\fr{dw_n}{dt}\right )^2
               +\fr{1}{2}K_2(w_n-v_n)^2+\fr{1}{2}K_{2}^{\prime}(v_{n+1}-w_n)^2
            \right.
               \nonumber\\
& & \hspace{1.65cm}\left.
           +\fr{1}{3}K_3(w_n-v_n)^3+\fr{1}{3}K_3^{\prime}(v_{n+1}-w_n)^3
               +\fr{1}{3}V_3v_n^3+\fr{1}{3}V_3^{\prime}w_n^3
        \right],
\eea
where $v_n$=$v_n(t)$\,($w_n=w_n(t)$\,) is the displacement from its equilibrium position
of the $n$th particle with mass $m$\,($M$). $n$ is the index of the $n$th 
unit cell with a lattice constant $a=2a_0$, $a_0$ is the equilibrium lattice 
spacing between two adjacent particles. 
Here for generality we assume that the nearest-neighbor
force constants $K_j(j=2,3)$ in the same cells 
are different from the nearest-neighbor force constants $K_j^{\prime}(j=2,3)$ 
in different cells. $V_3$ and $V_3^{\prime}$ are the force constants related
to the on-site cubic potential for two types of 
particles. Without loss of generality we assume $m<M$, 
$K_j^{\prime}\leq K_j(j=2,3)$, and $V_3^{\prime}\leq V_3$.
The equations of motion for describing the lattice read
\bea
&  &  \fr{d^2}{dt^2}v_n=
      I_2(w_n-v_n)+I_2^{\prime}(w_{n-1}-v_n)
      +I_3(w_n-v_n)^2-I_3^{\prime}(w_{n-1}-v_n)^2-\alpha_m v_n^2,\\
&  &  \fr{d^2}{dt^2}w_n=
      J_2(v_n-w_n)+J_2^{\prime}(v_{n+1}-w_n)
      -J_3(v_n-w_n)^2+J_3^{\prime}(v_{n+1}-w_n)^2-\alpha_M w_n^2,
\eea
where $I_j=K_j/m, I_j^{\prime}=K_j^{\prime}/m, J_j=K_j/M, 
J_j^{\prime}=K_j^{\prime}/M\,(j=2,3), \alpha_m=V_3/m$ and 
$\alpha_M=V_3^{\prime}/M$.  
The linear dispersion relation of Eqs.\,(2) and (3) is given by
\be
\omega_{\pm}(q)=\fr{1}{\sqrt{2}}\left\{ (I_2+I_2^{\prime}+J_2+J_2^{\prime})
\pm\left [ (I_2+I_2^{\prime}+J_2+J_2^{\prime})^2
-16 I_2 J_2^{\prime}\sin^2(qa/2)\right ]^{\fr{1}{2}}\right \}^{\frac{1}{2}},
\ee
where the minus\,(plus) sign corresponds to acoustic\,(optical) mode. Thus we 
have two phonon bands $\omega_{\pm}(q)$
and obviously $\om_{\pm}(q+Q)=\om_{\pm}(q)$,
here $Q=2j\pi/a$, $j$ is an integer and $Q$ is the reciprocal
lattice vector of the system. At the wave number $q=0$, 
the phonon spectrum has a lower cutoff
$\om_{-}(0)=0$ for the acoustic mode and an upper cutoff 
$\om_{+}(0)=(I_2+I_2^{\prime}+J_2+J_2^{\prime})^{1/2}$ for the optical mode. 
At $q=\pi/a$ there exists a frequency gap between the upper cutoff of the acoustic
branch $\om_{-}(\pi/a)$ and the lower cutoff of the optical branch 
$\om_{+}(\pi/a)$, where 
$\om_{\pm}(\pi/a)=(1/\sqrt{2})\{(I_2+I_2^{\prime}+J_2+J_2^{\prime})\pm
[(I_2+I_2^{\prime}+J_2+J_2^{\prime})^2-16 I_2 J_2^{\prime} ]^{1/2} \}^{1/2}$.
The width of the frequency gap is 
$\om_{+}(\pi/a)-\om_{-}(\pi/a)$, which approaches zero when
$m\rightarrow M$ and $K_2^{\prime}\rightarrow K_2$. This is just the limit
of monatomic lattice with the lattice constant $a_0=a/2$. We assume 
the gap is not small, i.\,e. we have 
$(1-m/M)$ and $(1-K_2/K_2^{\prime})$ are of order unity.

Because of the periodic property of $\om_{\pm}(q)$,  the condition of a
second harmonic resonance in the system (2) and (3) reads
\bea
& & q_2=2q_1+Q,\\
& & \om_2=2\om_1,
\eea
where $q_1\,(q_2)$ and $\om_1\,(\om_2)$  are the wave vector and frequency of
the corresponding fundamental\,(second harmonic) wave, respectively. 
Eqs.(5) and (6) 
are also called the phase-matching conditions for the SHG. It is easy to show that
in the limit $m\rightarrow M$ and $K_2^{\prime}\rightarrow K_2$ the 
condition (5) and (6) can not be satisfied except for zero-frequency mode,
i.\,e. the SHG is impossible in monatomic lattices. For the diatomic lattice,
in order to fulfil (5) and (6) we may chose $\om_1\in \om_{-}(q)$ and 
$\om_2\in \om_{+}(q)$, then the conditions (5) and (6) give
\bea
& & \left[ (I_2+I_2^{\prime}+J_2+J_2^{\prime})^2-4I_2 J_2^{\prime}
    \sin(q_1 a)\right]^{1/2}\nonumber\\
& & \hspace{0.8mm}=3(I_2+I_2^{\prime}+J_2+J_2^{\prime})
    -4\left[ (I_2+I_2^{\prime}+J_2+J_2^{\prime})^2
            -16 I_2 J_2^{\prime} \sin^2(q_1 a/2)\right]^{1/2}.
\eea
It is available to solve $q_1$ from the above equation. For simplicity we
consider cutoff modes of the system. We take $q_1=\pi/a, q_2=0$ and $Q=-2\pi/a$,
then the condition (5) is automatically satisfied. The condition (6)\,(the
same as (7)\,) now reads
\be
I_2+I_2^{\prime}+J_2+J_2^{\prime}=\fr{8}{\sqrt{3}} \sqrt{I_2 J_2^{\prime}}.
\ee
Eq.(8) also means that $\om_1=\om_{-}(\pi/a)=
(1/2)(I_2+I_2^{\prime}+J_2+J_2^{\prime})^{1/2}=(4 I_2 J_2^{\prime}/3)^{1/4}$
and 
$\om_2=\om_{+}(0)=
(I_2+I_2^{\prime}+J_2+J_2^{\prime})^{1/2}=2(4 I_2 J_2^{\prime}/3)^{1/4}$
If the all harmonic force constants are equal, i.\,e. $K_2^{\prime}=K_2$, 
Eq.(8) gives $m=M/3$. Another particular case is all masses are the same,
i.\,e. $m=M$.
In this case Eq.(8) requires $K_2^{\prime}=K_2/3$. In general, 
the phase-matching conditions (5) and (6) 
impose a constraint on masses and harmonic force constants of the lattice.
%
\subsection{Asymptotic expansion}
We employ the quasi-discreteness approach\,(QDA) developed in Refs.\,17 and 24 
for diatomic lattices to investigate the SHG in the system (2) and (3). 
We are interested in the cascading processes of the system in which
the width of excitation is narrower than usual SHG case. Thus
we use different spatial-temporal scales in deriving the envelope equations
for the fundamental and the second harmonic waves.  We make the expansion
\be
u_n(t)=\ep \left[ u^{(0)}_{n,n}+\ep^{1/2} u^{(1)}_{n,n}
             +\ep u^{(2)}_{n,n}+\cdots\right],
\ee
where $u_n(t)$ represents $v_n(t)$ or $w_n(t)$.
$\ep$ is a  smallness and ordering parameter denoting the relative 
amplitude of the excitation and
$u_{n,n}^{(\nu)}=u^{(\nu)}(\xi_n,\tau;\phi_n(t))$, with
\bea
& & \xi_n=\ep^{1/2} (na-\la t),\\
& & \tau=\ep t,\\ 
& & \phi_n=qna-\om (q)t,
\eea
where $\la$ is a parameter to be determined by a solvability 
condition\,(see below). Substituting (9)-(12) into Eqs.(2) and (3) and equating
the coefficients of the same powers of $\ep$, we obtain
\bea
&  &  \fr{\pa^2}{\pa t^2}v_{n,n}^{(j)}-I_2(w_{n,n}^{(j)}-v_{n,n}^{(j)})
      -I_2^{\prime}(w_{n,n-1}^{(j)}-v_{n,n}^{(j)})
      =M_{n,n}^{(j)},\\
&  &  M_{n,n}^{(0)}=0,\\
&  &  M_{n,n}^{(1)}=2\la \fr{\pa^2}{\pa t\pa \xi_n}v_{n,n}^{(0)}
      -I_2^{\prime} a \fr{\pa}{\pa \xi_n}w_{n,n-1}^{(0)},\\
&  &  M_{n,n}^{(2)}=2\la \fr{\pa^2}{\pa t\pa \xi_n}v_{n,n}^{(1)}
      -\left (2\fr{\pa^2}{\pa t\pa\tau}+\la^2 \fr{\pa^2}{\pa\xi_{n}^2}\right )
      v_{n,n}^{(0)}
      +I_2^{\prime} \left (  -a\fr{\pa}{\pa \xi_n} w_{n,n-1}^{(1)}
      +\fr{a^2}{2!}\fr{\pa^2}{\pa \xi_{n}^2}w_{n,n-1}^{(0)}\right )\nonumber\\
&  &  \hspace{1.5cm}+I_3(w_{n,n}^{(0)}-v_{n,n}^{(0)})^2
      -I_3^{\prime}(w_{n,n-1}^{(0)}-v_{n,n}^{(0)})^2-\al_m (v_{n,n}^{(0)})^2,\\
&  &  \vdots\hspace{1cm}\vdots \nonumber
\eea
and
\bea
&  &  \fr{\pa^2}{\pa t^2}w_{n,n}^{(j)}-J_2(v_{n,n}^{(j)}-w_{n,n}^{(j)})
      -J_2^{\prime}(v_{n,n+1}^{(j)}-w_{n,n}^{(j)})
      =N_{n,n}^{(j)},\\
&  &  N_{n,n}^{(0)}=0,\\
&  &  N_{n,n}^{(1)}=2\la \fr{\pa^2}{\pa t\pa \xi_n}w_{n,n}^{(0)}
      +J_2^{\prime} a\fr{\pa}{\pa \xi_n}v_{n,n+1}^{(0)},\\
&  &  N_{n,n}^{(2)}=2\la \fr{\pa^2}{\pa t\pa \xi_n}w_{n,n}^{(1)}
      -\left (2\fr{\pa^2}{\pa t\pa\tau}+\la^2 \fr{\pa^2}{\pa\xi_{n}^2}\right )
      w_{n,n}^{(0)}+J_2^{\prime}\left ( a\fr{\pa}{\pa \xi_n} v_{n,n+1}^{(1)}
      +\fr{a^2}{2!}\fr{\pa^2}{\pa \xi_{n}^2}v_{n,n+1}^{(0)}\right )\nonumber\\
&  &  \hspace{1.5cm}-J_3(v_{n,n}^{(0)}-w_{n,n}^{(0)})^2
      +J_3^{\prime}(v_{n,n+1}^{(0)}-w_{n,n}^{(0)})^2-\al_M (w_{n,n}^{(0)})^2,\\
&  &  \vdots\hspace{1cm}\vdots \nonumber
\eea
with $j=0,\,1,\,2, \cdots$. Eqs.(13) and (17) can be rewritten in the following 
form 
\bea
&   &  \hat{L}w_{n,n}^{(j)}=J_2 M_{n,n}^{(j)}+J_2^{\prime}M_{n,n+1}^{(j)} 
       +\left (\fr{\pa^2}{\pa t^2}+I_2+I_2^{\prime}\right )N_{n,n}^{(j)},\\
&   &  \left (\fr{\pa^2}{\pa t^2}+I_2+I_2^{\prime}\right )v_{n,n}^{(j)}=I_2
       w_{n,n}^{(j)}+I_2^{\prime}w_{n,n-1}^{(j)}+M_{n,n}^{(j)},
\eea
where the operator $\hat{L}$ is defined by
\bea
& & \hat{L}u_{n,n}^{(j)}=\left(\fr{\pa^2}{\pa t^2}+I_2+I_2^{\prime}\right )
                     \left(\fr{\pa^2}{\pa t^2}+J_2+J_2^{\prime}\right )
                     u_{n,n}^{(j)}
                     -(I_2 J_2+I_2^{\prime} J_2^{\prime})u_{n,n}^{(j)}
                     \nonumber\\
& & \hspace{1.6cm}-I_2 J_2^{\prime}
                     \left(u_{n, n+1}^{(j)}+u_{n,n-1}^{(j)}\right),
\eea
where $u_{n,n}^{(j)}(j=0,\,1,\,2,\cdots)$ are a  set of arbitrary functions.
From Eq.(21) we can solve $w_{n,n}^{(j)}$ and obtain a series of 
solvability conditions\,(envelope equations) whereas Eq.(22) is used to solve 
$v_{n,n}^{(j)}$.
%
\subsection{Envelope equations for cascading processes}
We now solve Eqs.(22) and (23) order by order. 
For $j=0$ it is easy to get
\bea
&  &  w_{n,n}^{(0)}=A_{1}(\tau,\xi_n)\exp(i\phi_{n}^{-})
                    +A_{2}(\tau,\xi_n)\exp(i\phi_{n}^{+})
                    +{\rm c.\,c.},\\
&  &  v_{n,n}^{(0)}=\fr{ I_2+I_2^{\prime} e^{-iqa} }{-\om_{-}^{2}
                    +I_2+I_2^{\prime} } A_{1}(\tau,\xi_n)
                     \exp(i\phi_{n}^{-})
                    +\fr{ I_2+I_2^{\prime} e^{-iqa} }{-\om_{+}^{2}
                    +I_2+I_2^{\prime} } A_{2}(\tau,\xi_n)
                     \exp(i\phi_{n}^{+})
                    +{\rm c.\,c.}
\eea
with $\phi_n^{\pm}=qna-\om_{\pm}(q) t$. $\om_{\pm}(q)$ have been given in Eq.(4).
$A_1$ and $A_2$ are yet to be determined two envelope\,(or amplitude) functions
of the acoustic and the optical excitations, respectively. They are the
functions of the slow variables $\xi_n$ and $\tau$. c.\,c. denotes the corresponding
complex conjugate. For simplicity we specify two modes, i.\,e. the
acoustic upper cutoff mode ($q_1=\pi/a$, 
$\om_1=\om_{-}(\pi/a)=(4I_2 J_2^{\prime}/3)^{1/4}$\,) and the optical
upper cutoff mode
($q_2=0$, $\om_2=\om_{+}(0)=2\om_1=2(4I_2 J_2^{\prime}/3)^{1/4}$\,). Thus we have 
\bea
&  &  w_{n,n}^{(0)}=A_{1}(\tau,\xi_n) (-1)^n \exp(-i \om_1 t)
                    +A_{2}(\tau,\xi_n)\exp(-i \om_2 t)
                    +{\rm c.\,c.},\\
&  &  v_{n,n}^{(0)}=\fr{ I_2-I_2^{\prime}  }{-\om_{1}^{2}
                    +I_2+I_2^{\prime} } A_{1}(\tau,\xi_n) (-1)^n
                     \exp(-i\om_1 t)\nonumber\\
&  & \hspace{1.2cm}+\fr{ I_2+I_2^{\prime} }{-\om_{2}^{2}
                    +I_2+I_2^{\prime} } A_{2}(\tau,\xi_n)
                     \exp(-i\om_2 t)
                    +{\rm c.\,c.}.
\eea
From the discussion in subsection II.\,A, the modes chosed in such way
satisfy the phase-matching conditions (5) and (6) for the SHG. Thus in Eqs.(27)
and (28) $A_1$\,($A_2$) represents the amplitude of the 
fundamental\,(second harmonic) wave, respectively.

In the next order ($j$=1), a solvability condition of Eqs.(21) and (22) requires
the parameter $\la=0$, thus $\xi_n=na$. The second-order solution reads
\bea
& & w_{n,n}^{(1)}=B_0+\left[ B_1 (-1)^n \exp (-i \om_1 t)
               +B_2 \exp (-i \om_2 t)+{\rm c.\,c.}  \right],\\
& & v_{n,n}^{(1)}=B_0+\left\{ 
                \fr{ (I_2-I_2^{\prime})B_1+I_2^{\prime} a \pa A_1/\pa \xi_n }
                   {-\om_1^2+I_2+I_2^{\prime} }
                   (-1)^n \exp (-i \om_1 t)\right .\nonumber\\
& & \hspace{2.2cm}+\left. \fr{ (I_2+I_2^{\prime})B_2-I_2^{\prime} a \pa A_2/\pa \xi_n }
                   {-\om_2^2+I_2+I_2^{\prime} }
                \exp (-i \om_2 t)+{\rm c.\,c.}\right\},
\eea
where $B_j$\,($j$=0,\,1,\,2) are undetermined functions of $\xi_n$ and $\tau$.

In the order $j$=2, we have the third-order approximate equatuion
\be
\hat{L}w_{n,n}^{(2)}=J_2 M_{n,n}^{(2)}+J_2^{\prime}M_{n,n+1}^{(2)} 
       +\left (\fr{\pa^2}{\pa t^2}+I_2+I_2^{\prime}\right )N_{n,n}^{(2)}.
\ee
Eq.(22) is not necessary since from (30) we can obtain closed equations 
for $A_1$ and $A_2$. Using Eqs.(26)-(29) we can get $M_{n,n}^{(2)}$,
$M_{n,n+1}^{(2)}$ and $N_{n,n}^{(2)}$. By a detailed calculation 
we obtain the sovability condition of Eq.(30) 
\bea
&  &  i\fr{\pa A_1}{\pa \tau}+\fr{1}{2}\Gamma_{1}\fr{\pa^2 A_1}{\pa \xi_{n}^{2}}
      +\Delta_{1}\,A_{1}^{*} A_{2}=0,\\
&  &  i\fr{\pa A_1}{\pa \tau}+\fr{1}{2}\Gamma_{2}\fr{\pa^2 A_2}{\pa \xi_{n}^{2}}
      +\Delta_{2}\,A_{1}^{2} =0,
\eea
where the coefficients are expressed as
\bea
& & \Gamma_1=-\fr{I_2^{\prime} J_2^{\prime} a^2}
                 {\om_1 [\la_1^{-1}+\la_1 (I_2-I_2^{\prime})(J_2-J_2^{\prime})]},\\
& & \Gamma_2=-\fr{I_2 J_2^{\prime} a^2}
                 {\om_2 [-\la_2^{-1}-\la_2 (I_2+I_2^{\prime})(J_2+J_2^{\prime})] },\\
& & \Delta_1=\fr{ [1-\la_2(I_2+I_2^{\prime})]\la_3-\la_1^{-1}\al_M
                  -\la_1 \la_2 (I_2^2-(I_2^{\prime})^2)(J_2-J_2^{\prime})\al_m 
                }
                {\om_1 [\la_1^{-1}+\la_1 (I_2-I_2^{\prime})(J_2-J_2^{\prime})]},\\ 
& & \Delta_2=\fr{ \la_4-\la_2^{-1}\al_M
                  -\la_1^2 (I_2-I_2^{\prime})^2 (J_2+J_2^{\prime})\al_M }
                {2\om_2 [\la_2^{-1}+\la_2 (I_2+I_2^{\prime})(J_2+J_2^{\prime})]},   
\eea
with
\bea
& & \la_j=\fr{1}{-\om_j^2+I_2+I_2^{\prime}},\hspace{1.2cm}(j=1,2)\\
& & \la_3=(I_3-I_3^{\prime})(J_2-J_2^{\prime})-\la_1^{-1}(J_3-J_3^{\prime})
          +(I_2-I_2^{\prime}) \left [(J_3+J_3^{\prime}) -\la_1 (I_3+I_3^{\prime})
          (J_2-J_2^{\prime})\right ],\\
& & \la_4=[1-\la_1(I_2-I_2^{\prime})]^2[-J_3\la_2^{-1}+I_3(J_2+J_2^{\prime})]
          +[1+\la_1(I_2-I_2^{\prime})]^2[J_3^{\prime}\la_2^{-1}-I_3
          ^{\prime}(J_2+J_2^{\prime})].
\eea    
Introducing the transformation $u_j=\ep A_j\,(j=1,2)$ and noting that 
$\xi_n=\ep^{1/2} x_n\,(x_n\equiv na)$ and $\tau=\ep t$, Eqs.(31) and (32)
can be rewritten into the form
\bea
&  &  i\fr{\pa u_1}{\pa t}+\fr{1}{2}\Gamma_{1}\fr{\pa^2 u_1}{\pa x_{n}^{2}}
      +\Delta_{1}\,u_{1}^{*} u_{2}=0,\\
&  &  i\fr{\pa u_2}{\pa t}+\fr{1}{2}\Gamma_{2}\fr{\pa^2 u_2}{\pa x_{n}^{2}}
      +\Delta_{2}\,u_{1}^{2} =0.
\eea
We should point out that Eqs.(5) and (6) are perfect phase-matching conditions
for the SHG. If we allow a small mismatch for frequency, $\delta \om$,
the conditions (5) and (6) become
\be
\om_2=2\om_1+\delta \om, \hspace{1cm} q_2=2q_1+Q.
\ee
In this case Eqs.(40) and (41) change into 
\bea
&  &  i\left( \fr{\pa u_1}{\pa t}+v_1 \fr{\pa u_1}{\pa x_n}\right)
      +\fr{1}{2}\Gamma_{1}\fr{\pa^2 u_1}{\pa x_{n}^{2}}
      +\Delta_{1}\,u_{1}^{*} u_{2}\exp (-i \delta\om t)=0,\\
&  &  i\left( \fr{\pa u_2}{\pa t}+v_2 \fr{\pa u_2}{\pa x_n}\right)
      +\fr{1}{2}\Gamma_{2}\fr{\pa^2 u_2}{\pa x_{n}^{2}}
      +\Delta_{2}\,u_{1}^{2}\exp(i \delta\om t) =0,
\eea
where $v_j\,(j=1,2)$ are the group velocities of of the fundamental and the
second harmonic waves near at $q=\pi/a$ and $q=0$, respectively.

Eqs.(43) and (44) are the coupled-mode equations for the fundamental and
the second harmonic waves. Such equations have been obtained by Karamzin
and Sukhorukov in the context of nonlinear optics[35]. One of important features
of the KS equations is the inclusion of dispersion, which is absent in usual 
SHG envelope equations[41].
%
%
\section{LATTICE SIMULTON SOLUTIONS}
In this section, we solve the KS equations (43) and (44) derived in our lattice
model and thus present some lattice simulton solutions for the system (2) and (3).
In general, the property of the solutions of Eqs.(43) and (44) depends strongly
on the coefficients appearing in the equations, in particular on their signs. At
first we notice that in our system, $\Gamma_1$ and $\Gamma_2$, which are 
respectively the group-velocity dispersion of the fundamental and the second 
harmonic waves, are both negative. But the signs of the nonlinear coefficients,
$\Delta_1$ and $\Delta_2$, may be  generally of both signs.  Thus
the situation here is different from the KS equations derived for the cascading
process in nonlinear optics, where the nonlinear coefficients have the same sign,
while the group-velocity dispersions may have different signs[42].

To solve Eqs.(43) and (44), we make the transformation
\bea
& & u_1(x_n,t)=U_1(\zeta) \exp[i(k_1x_n-\Omega_1 t)],\\
& & u_2(x_n,t)=U_2(\zeta) \exp[i(k_2x_n-\Omega_2 t)],
\eea
with $\zeta=k x_n-\Omega t$. Substituting (45)
and (46) into (43) and (44), we obtain
\bea
& & \fr{d^2 U_1}{d\zeta^2}+\al_1 U_1 U_2-\beta_1 U_1=0,\\
& & \fr{d^2 U_2}{d\zeta^2}+\al_2 U_1^2-\beta_2 U_2=0,
\eea
where $\al_1=2\Delta_1/(\Gamma_1 k^2), \al_2=2\Delta_2/(\Gamma_2 k^2),
\beta_1=-2(\Omega_1-v_1k-\fr{1}{2}\Gamma_1 k_1^2)/(\Gamma_1 k^2),
\beta_2=-2(\Omega_2-v_2k-\fr{1}{2}\Gamma_2 k_2^2)/(\Gamma_2 k^2)$,
$\Omega=v_1k+\Gamma_1kk_1$, with $k_2=2k_1, \Omega_2=2\Omega_1+\delta\om$
and $k_1=(v_2-v_1)/(\Gamma_1-2\Gamma_2)$. 
One of the coupled soliton-soliton\,(i.\,e. simultaneous solitons for two wave
components)  solutions of Eqs.(47) and (48) reads
\bea
& & U_1=\fr{6}{\sqrt{\al_1 \al_2}} \left( \fr{2}{3}-{\rm sech}^2\zeta\right ),\\
& & U_2=-\fr{6}{\al_1} \left( \fr{2}{3}-{\rm sech}^2\zeta\right ),
\eea
where a condition $\beta_1=\beta_2=-4$ is required. The parameter $k$ is given by
\be
k=\fr{ 2(v_2-v_1)k_1+(\Gamma_1-2\Gamma_2)k_1^2+\delta\om  }
     {2(\Gamma_2-2\Gamma_1)}.
\ee
The lattice configuration in this case takes the form
\bea
& & w_n(t)=(-1)^n \fr{12}{\sqrt{\al_1\al_2}}
           \left[ \fr{2}{3}-{\rm sech}^2(kna-\Omega t)\right] 
           \cos[k_1na-(\om_1+\Omega_1)t]\nonumber\\
& & \hspace{1.5cm}-\fr{12}{\al_1}
           \left[\fr{2}{3}-{\rm sech}^2(kna-\Omega t)\right] 
           \cos[k_2na-(\om_2+\Omega_2)t],\\
& & v_n(t)=(-1)^n \fr{12}{\sqrt{\al_1\al_2}}
           \fr{ I_2-I_2^{\prime} }{ -\om_1^2+I_2+I_2^{\prime} }
           \left[ \fr{2}{3}-{\rm sech}^2(kna-\Omega t)\right] 
           \cos[k_1na-(\om_1+\Omega_1)t]\nonumber\\
& & \hspace{1.5cm}-\fr{12}{\al_1}
           \fr{ I_2+I_2^{\prime} }{ -\om_2^2+I_2+I_2^{\prime} }
           \left[\fr{2}{3}-{\rm sech}^2(kna-\Omega t)\right] 
           \cos[k_2na-(\om_2+\Omega_2)t]. 
\eea
If $q_1\,(q_2)$ is exactly equal to $\pi/a$\,(zero) but with $\delta \om\neq 0$,
one has $v_1=v_2=0$. In this case $k_1=k_2=0$, $\Omega_1=2\Gamma_1k^2, 
\Omega_2=2\Gamma_2k^2, \Omega=0$ and 
$k=\{\delta\om/[2(\Gamma_2-2\Gamma_1)]\}^{1/2}$.
(52) and (53) present a {\it nonpropagating simulton} excitation, in which
the vibrating frequency of the acoustic-\,(optical-) mode component being within
the acoustic(optical) phonon band. In our model, $\Gamma_2-2\Gamma_1>0$ thus
$\delta \om$ should be taken positive in this case. In addition, from
(52) and (53) we see that the envelopes for both the acoustic and optical 
components are kinks (or dark solitons). Futhermore,
if $K_2^{\prime}=K_2$, the displacement of light particles, $v_n(t)$,
only has an optical-mode component.

The other simulton solution of Eqs.(47) and (48) reads
\bea
& & U_1=-\fr{6}{\sqrt{\al_1 \al_2}} {\rm sech}^2\zeta,\\
& & U_2=-\fr{6}{\al_1} \left( \fr{4}{3}-{\rm sech}^2\zeta\right ),
\eea
where we have $\beta_1=-\beta_2=-4$. The parameter $k$ now reads
\be
k=\fr{ 2(v_2-v_1)k_1+2(\Gamma_2-\Gamma_1)k_1^2-\delta\om  }
     {2(2\Gamma_1+\Gamma_2)}.
\ee
The lattice configuration is now given by
\bea
& & w_n(t)=(-1)^{n+1} \fr{12}{\sqrt{\al_1\al_2}}
           {\rm sech}^2(kna-\Omega t) 
           \cos[k_1na-(\om_1+\Omega_1)t]\nonumber\\
& & \hspace{1.5cm}-\fr{12}{\al_1}
           \left(\fr{4}{3}-{\rm sech}^2(kna-\Omega t)\right) 
           \cos[k_2na-(\om_2+\Omega_2)t],\\
& & v_n(t)=(-1)^{n+1} \fr{12}{\sqrt{\al_1\al_2}}
           \fr{ I_2-I_2^{\prime} }{ -\om_1^2+I_2+I_2^{\prime} }
           {\rm sech}^2(kna-\Omega t) 
           \cos[k_1na-(\om_1+\Omega_1)t]\nonumber\\
& & \hspace{1.5cm}-\fr{12}{\al_1}
           \fr{ I_2+I_2^{\prime} }{ -\om_2^2+I_2+I_2^{\prime} }
           \left[\fr{4}{3}-{\rm sech}^2(kna-\Omega t)\right] 
           \cos[k_2na-(\om_2+\Omega_2)t]. 
\eea                
Thus in this case the acoustic-mode component is a staggered envelope soliton
but the optical-mode component is still an envelope kink. 
If $v_1=v_2=0$ we have $k_1=k_2=0, \Omega_1=2\Gamma_1 k^2, \Omega_2=-2\Gamma_2 k^2,
\Omega=0$ and $k=\{-\delta\om/[2(\Gamma_2-2\Gamma_1)]\}^{1/2}$. 
In this situation the simulton (57) and (58) is also a nonpropagating excitation
with the vibrating frequency of the acoustic-\,(optical-)  mode component 
within(above) the acoustic(optical) phonon band. 
In order to make $k$ to be real we should take $\delta \om<0$ in this case.

A common requirement for the existence of the simulton 
solutions (52), (53), (57) and
(58) is sgn($\al_1 \al_2)>0$, which means sgn($\Delta_1 \Delta_2)>0$
because $\Gamma_1\Gamma_2>0$ in our model.  It can be met by chosing 
different values of system parameters. For example, in the following
two particular cases we have sgn($\Delta_1 \Delta_2)>0$:
\begin{enumerate}
  \item $K_2^{\prime}=K_2, K_3^{\prime}=K_3=0$. In this case 
        $\Delta_1=-\al_M/\om_1, \Delta_2=-J_2\al_M/[2\om_2(I_2+J_2)]$.
  \item $K_2^{\prime}=K_2, V_3^{\prime}=V_3=0$. In this case 
        $\Delta_1=(J_3^{\prime}-J_3)(1+I_2/J_2)/\om_1$,
        $\Delta_2=(I_3^{\prime}-I_3+J_3^{\prime}-J_3)/[2\om_2 (I_2+J_2)]$.
\end{enumerate}
%
%
\section{DISCUSSION AND SUMMARY}
We have {\it analytically} shown that the lattice simultons are possible 
in nonlinear diatomic lattices. Based on the QDA for the nonlinear excitations
in diatomic lattices developed before[17,24], we have considered the resonant
coupling between two phonon modes, one from the acoustic and other one from
the optical branch, respectively. The KS equations are derived for the envelopes
of the fundamental and second harmonic waves by taking new multiple spatial-temporal
scale variables, which are necessary for narrower nonlinear excitations. 
Exact coupled soliton\,(simulton) solutions are obtained for 
the KS equations and the simulton configurations for the lattice displacements are 
explicitly given.

Similar to the optical simultons in nonlinear optical media, the physical
mechanism for the formation of the lattice simultons is due to the cascading
effect between two lattice wave components. In this process, the fundamental
and the second harmonic waves interact with themselves through repeated
wave-wave interactions. For instance the energy of the fundamental wave
is first upconverted to the second harmonic wave and then downconverted
again, resulting in a mutual self-trapping of each wave thus the formation of 
two simultaneous solitons.

Mathematically, in addition to the resonance conditions (5) and (6), the 
formation of a lattice simulton needs a balance between the cubic 
nonlinearity\,(in the Hamiltonian) and the dispersion,  the latter
is provided by the discreteness of the system. Thus for deriving the
envelope equations in this case, we must chose the multiple-scale
variables different from the ones used in usual SHG. In our derivation for
the KS equations based on the QDA[17,24], only one small parameter,
i.\,e. the amplitude of the excitation, is used. This method gives 
a clear, justified and self-consistent hierarchy of scales and thus the
corresponding solvability conditions, which are just the envelope 
equations we need. Thus it is satisfactory according to  the point of 
view of singular perturbation theory.

Cubic nonlinearity exists in most of realistic atomic potentials[24]. Thus
it is possible to observe the lattice simultons reported here. It must be 
emphasized that the multi-value property of the linear dispersion relation
is important for generating the simultons in lattices. Thus a diatomic or 
multi-atomic lattice is necessary for observing such excitations.

The theory given above can be applied to multi-atomic and higher-dimensional
lattices, and higher-order nonlinearity can also be included. For instance,
if we consider the Hamiltonian with cubic and quartic nonlinearities,
Eqs.(31) and (32) sould be generalized to
\bea
&  &  i\fr{\pa A_1}{\pa \tau}+\fr{1}{2}\Gamma_{1}\fr{\pa^2 A_1}{\pa \xi_{n}^{2}}
      +\Delta_{1}\,A_{1}^{*} A_{2}
      +(\Lambda_{11}|A_1|^2+\Lambda_{12}|A_2|^2)A_1=0,\\
&  &  i\fr{\pa A_1}{\pa \tau}+\fr{1}{2}\Gamma_{2}\fr{\pa^2 A_2}{\pa \xi_{n}^{2}}
      +\Delta_{2}\,A_{1}^{2}
      +(\Lambda_{21}|A_1|^2+\Lambda_{22}|A_2|^2)A_2=0,
\eea
where $\Lambda_{ij}(i,j=1,2)$ are self-phase and cross-phase modulational
coefficients contributed by the quartic nonlinearity of the system.
Eqs.(59) and (60) can be derived using the multiple-scale variables
$\xi_n=\ep x_n, \tau=\ep^2 t$ under the assumption 
$v_n(t)=O(\ep), w_n(t)=O(\ep)$, $K_3=O(K_3^{\prime})=O(\ep)$, and
$V_3=O(V_3^{\prime})=O(\ep)$. A small frequency mismatch can also be included in 
(59) and (60) and similar equations like (43) and (44) with additional
self- and cross-phase modulational terms can also be written down.
A detailed study will be  presented elsewhere.
%
%
%
\section*{ACKNOWLEDGMENTS}
This work is supported in part by grants from the Hong Kong
Research Grants Council(RGC), the Hong Kong Baptist University Faculty Research
Grant(FRG), the Natural Science Foundation of China, and the Development
Foundation for Science and Technology of ECNU.

\end{document}